\documentclass[
  aip,
  jap,
  twoside,
  twocolumn,
  showpacs,
  floatfix,
  10pt,
]{revtex4-1}

\usepackage{amsmath}
\usepackage{amssymb}
\usepackage{graphicx}
\usepackage{tabularx}
\usepackage{dcolumn}

\newcommand{\Al}[0]{\text{Al}}
\newcommand{\Ti}[0]{\text{Ti}}
\newcommand{\Y}[0]{\text{Y}}
\newcommand{\He}[0]{\text{He}}

\renewcommand{\O}[0]{\text{O}}

\newcommand{\ion}[0]{\text{ion}}
\newcommand{\bulk}[0]{\text{bulk}}

\newcommand{\eV}[0]{\text{eV}}

\newcommand{\VBM}[0]{\text{VBM}}

\newcommand{\LDA}[0]{\text{LDA}}

\newcommand{\bcc}[0]{\text{bcc}}
\newcommand{\rocksalt}[0]{\text{rocksalt}}

\newcommand{\sect}[1]{Sect.~\ref{#1}}
\newcommand{\fig}[1]{Fig.~\ref{#1}}
\newcommand{\Eq}[1]{Eq.~(\ref{#1})}

\newcommand{\tab}[1]{Table~\ref{#1}}

\renewcommand{\epsilon}[0]{\varepsilon}

\begin{document}


\title{
  A first-principles study of helium storage in oxides and at oxide--iron interfaces
}

\author{Paul Erhart}
\email{erhart@chalmers.se}
\affiliation{
  Physical and Life Science Directorate,
  Lawrence Livermore National Laboratory,
  Livermore, California, USA
  and
  Department of Applied Physics,
  Chalmers University of Technology,
  Gothenburg, Sweden 
}

\begin{abstract}
Density-functional theory calculations based on conventional as well as hybrid exchange-correlation functionals have been carried out to study the properties of helium in various oxides (Al$_2$O$_3$, TiO$_2$, Y$_2$O$_3$, YAP, YAG, YAM, MgO, CaO, BaO, SrO) as well as at oxide-iron interfaces.
Helium interstitials in bulk oxides are shown to be energetically more favorable than substitutional helium, yet helium binds to existing vacancies. The solubility of He in oxides is systematically higher than in iron and scales with the free volume at the interstitial site nearly independently of the chemical composition of the oxide. In most oxides He migration is significantly slower and He--He binding is much weaker than in iron. To quantify the solubility of helium at oxide-iron interfaces two prototypical systems are considered (Fe\textbar MgO, Fe\textbar FeO\textbar MgO). In both cases the He solubility is markedly enhanced in the interface compared to either of the bulk phases. The results of the calculations allow to construct a schematic energy landscape for He interstitials in iron. The implications of these results are discussed in the context of helium sequestration in oxide dispersion strengthened steels, including the effects of interfaces and lattice strain.
\end{abstract}

\maketitle


\section{Introduction}

Fusion environments are characterized by an abundance of energetic helium ions ($\alpha$-particles, He$^{2+}$) that are produced in the fusion reaction as well as via nuclear transmutation reactions in the plasma facing components. \cite{Ehr99, BloBusDut07} Candidate materials for structural applications in fusion reactors must therefore be able to tolerate large concentrations of helium without degradation of their mechanical properties under extreme conditions of up to 200 displacements per atom and 2000 appm He. \cite{OdeAliWir08} Most conventional steels are unable to sustain these conditions since they suffer from He mediated bubble as well as void formation and growth (``swelling'') leading to embrittlement and mechanical failure. \cite{OdeAliWir08}

Using first-principles calculations it has been shown that in iron (as a model system for steels) He prefers to occupy substitutional rather than interstitial sites. \cite{SelOseSto05, FuWil05} For dynamical reasons, however, during irradiation the majority of He is introduced in the form of interstitials. \cite{ErhMar11} Since the effective interaction between He interstitials in the iron matrix is attractive \cite{FuWil05} they can readily form clusters. The latter are further stabilized by the addition of vacancies, which eventually leads to the formation and growth of bubbles and voids.

To overcome the shortcomings of conventional steels for applications in fusion environments, it has been suggested to employ nano-structured ferritic alloys, \cite{OdeAliWir08} in particular oxide dispersion strengthened (ODS) steels. These materials are obtained by mechanical alloying of oxide particles with steel powders (see e.g., Refs.~\onlinecite{UkaNisOka97, UkaNisOku98, KasTodYut07}). They are characterized by a fine distribution of nanometer-sized oxide particles \cite{FuKrcPai07, Mar08, JiaSmiOde09} that act as obstacles for dislocation motion and are metastable up to very high temperatures. The oxide particles and even more so the oxide-matrix interfaces are expected to act as sinks for He interstitials. The very high density of these particles should lead to a fine distribution of He bubbles and thereby effectively limit the formation of larger supercritical voids that lead to mechanical failure. \cite{OdeAliWir08} Several experiments provided evidence that ODS steels are much more swelling resistant than their non-oxide containing counterparts, raising hopes that these materials will eventually satisfy the design criteria for future fusion reactors.

Designing ODS steels that can sustain the extreme conditions inside a fusion reactor requires a close collaboration between experiment and modeling. As there are currently no neutron sources available that can reproduce the intense neutron spectrum resulting from fusion, a combination of (experimental and numerical) simulations must be employed to predict material performance. In this context, numerical modeling of the time evolution of defect populations (vacancies, interstitials, interstitial loops, dislocations, bubbles, voids \textit{etc.}) plays a pivotal role. \cite{Man87, MarBul11} Such models rely on a database of rates for various microscopic processes that occur in the material, which typically comprises both experimental and atomic scale modeling data. Given the importance of oxides particles in improving swelling resistance, the microscopic mechanisms that govern their interaction with He deserve particular attention, yet at present our microscopic understanding of these interactions is limited.

A recent very extensive transmission electron microscopy study \cite{HsiFluTum10} revealed that in ODS steels bubbles and voids form preferentially in the close vicinity of oxide particles, leading in many cases to the formation of a ``ring'' of bubbles surrounding a particle. The same study also demonstrated that oxide particles in ODS steels exhibit a broad variety of size, chemical composition, bulk and interface structure including amorphous, partially amorphous and crystalline particles, core-shell configurations as well as chemical gradients. It was furthermore shown that materials with smaller particles can tolerate more helium providing direct evidence for the importance of the interface area. \cite{HsiFluTum10} In this context, an oxide ``particle'' can be as small as a few {\AA}ngstr\"oms, correponding to just a couple of atoms. \cite{FuKrcPai07, JiaSmiOde09} While size of these ``nanofeatures'' \cite{OdeAliWir08} can be below the resolution obtainable in transmission electron microscopy, they nonetheless contribute to He sequestration due to their sheer number and very large effcetive interface area.

Atomic scale modeling is in principle ideally suited to complement these experiments and to provide detailed insight as well as quantitative information regarding the behavior of helium inside and near oxide particles. The enormous chemical and structural complexity present in ODS steels, however, renders a direct simulation of these systems at present impractical. On the other hand macroscopic measurements of He populations indicate that the most important parameter is particle size and thus interface area. \cite{HsiFluTum10} This suggests that the solubility of helium follows more general trends that are independent of variations in the local chemistry. The objective of the present paper is to demonstrate that the behavior of He in oxides and in the vicinity of oxide-iron interfaces can to a large extent be described by simple scaling relations. This is achieved by means of density-functional theory calculations using both conventional and hybrid exchange-correlation (XC) functionals of bulk oxides, bulk iron, and representative oxide-iron interfaces. The thus obtained data provides not only valuable insight into the microscopic mechanisms but can be further utilized to parametrize for example rate equation models. \cite{MarBul11} Note that even though nanosized oxide nuclei of the type described in Refs.~\onlinecite{FuKrcPai07} and \onlinecite{JiaSmiOde09} are not explicitly studied in the present work, the arguments that indicate that the governing parameter is free volume and in particular free volume at oxide-iron interfaces in general also transpire to the case of very small oxide inclusion. Compared to earlier investigations that considered He defects in select oxides and carbides, \cite{GryRasKot10, CheYinZha11, Yak11} the present work aims to provide a more general perspective, including a variety of oxides that represent different local environments, chemistry, and covalent-{\it vs}-ionic character.

The paper is organized as follows. In the next section computational methods and parameters are summarized. The results of a comprehensive study of He-related defects in three prototypical oxides (Al$_2$O$_3$, TiO$_2$, Y$_2$O$_3$) are described in \sect{sect:defects}. It is demonstrated that He occurs predominantly in the form of interstitials and that solubilities are much larger than in iron. Furthermore it is found that formation energies of He interstitials determined using conventional XC functionals are very close to values obtained from (more elaborate) hybrid XC functional calculations. To rationalize the variation of He interstitial formation energies among different oxides, in \sect{sect:interstitials} their volume dependence is studied additionally including the oxides Y$_4$Al$_2$O$_9$ (YAM), Y$_3$Al$_5$O$_{12}$ (YAG), YAlO$_3$ (YAP), MgO, CaO, SrO, and BaO. Section~\ref{sect:migbarrs} concerns the migration barriers for He interstitials in oxides, which are found to be systematically higher than in iron. The binding between He interstitials is the subject of \sect{sect:clustering}, where it is shown that most oxides are less prone to He cluster formation than Fe, a behavior that results from a lower density of interstitial site in these materials. The solubility of He interstitials in two representative oxide-iron interfaces (Fe\textbar MgO, Fe\textbar FeO\textbar MgO) is quantified in \sect{sect:interfaces}, where solubilities at interfaces are found to be systematically higher than in the bulk phases. Finally all these data are combined to sketch a typical energy landscape for He interstitial migration across an oxide-iron interface and discuss the results in the context of He sequestration in ODS steels.


\section{Methodology}

\begin{table*}
  \newcommand{\spr}[1]{\multicolumn{1}{c}{#1}}
  \centering
  \caption{
    Overview of computational parameters used in calculations of properties of the ideal bulk systems as well as point defects. Migration barrier calculations for Y$_2$O$_3$ and the rocksalt structured oxides were carried using the parameters given in brackets.
  }
  \label{tab:method}
  \begin{tabular}{ll*{7}c}
    \hline\hline
    &
    & Al$_2$O$_3$
    & TiO$_2$
    & Y$_2$O$_3$
    & YAP
    & YAG
    & YAM
    & (Mg,Ca,Ba,Sr)O \\
    \hline
    \multicolumn{7}{l}{Ideal cell calculations} \\
    & Number of atoms
    & 10 & 6 & 40 & 20 & 80 & 60 & 2
    \\
    & $k$-point sampling
    & $\Gamma 4\times 4\times 4$  
    & $\Gamma 6\times 6\times 6$  
    & $\Gamma 2\times 2\times 2$  
    & $\Gamma 4\times 4\times 4$  
    & $\Gamma 2\times 2\times 2$  
    & $\Gamma 2\times 2\times 2$  
    & $\Gamma 8\times 8\times 8$  
    \\
    \hline
    \multicolumn{7}{l}{Defect calculations} \\
    & Number of atoms
    & 270 & 216 & 320$^a$ & 160 & 160 & 60 & 216$^b$
    \\
    & Type of supercell
    & rhombohedral & tetragonal & body-centered cubic
    & orthorhombic & simple cubic & monoclinic & simple cubic
    \\
    & Supercell size
    & $3\times 3\times 3$  
    & $3\times 3\times 4$  
    & $2\times 2\times 2$  
    & $2\times 2\times 2$  
    & $1\times 1\times 1$  
    & $1\times 1\times 1$  
    & $3\times 3\times 3$  
    \\
    & $k$-point sampling
    & $\Gamma 1\times 1\times 1$  
    & $\Gamma 1\times 1\times 1$  
    & $\Gamma 1\times 1\times 1$  
    & $\Gamma 1\times 1\times 1$  
    & $\Gamma 1\times 1\times 1$  
    & $\Gamma 1\times 1\times 1$  
    & $\Gamma 1\times 1\times 1$  
    \\
    & XC functional
    & PBE, HSE06  
    & LDA, HSE06  
    & PBE  
    & PBE  
    & PBE  
    & PBE  
    & PBE  
    \\
    \hline\hline
    \multicolumn{9}{l}{
      $^a$ Migration barriers calculated using 80-atom cells and a $\Gamma 2\times 2\times 2$ $k$-point sampling.}
    \\
    \multicolumn{9}{l}{
      $^b$ Migration barriers calculated using 64-atom cells and a $\Gamma 4\times 4\times 4$ $k$-point sampling.}
  \end{tabular}
\end{table*}

\subsection{Thermodynamics}

Whereas in elemental metals the formation energy of an intrinsic defect is constant, in oxides defect formation energies depend both on the electron chemical potential (also referred to as Fermi level) and the chemical environment. The expression for the defect formation energy makes these dependencies explicit, \cite{ZhaNor91, ErhAbeLor10}
\begin{align}
  \Delta E_D
  &= (E_D - E_H) - q (E_{\VBM} + \mu_e) - \sum_i \Delta n_i \mu_i,
  \label{eq:eform}
\end{align}
where $\mu_i$ is the chemical potential of component $i$, $E_D$ is the total energy of the system containing the defect and $E_H$ is the total energy of the ideal reference system. The defect formation energy is linearly dependent on the defect charge state $q$ and the electron chemical potential $\mu_e$ which is measured with respect to the valence band maximum $E_{\VBM}$. Here, $\Delta n_i$ denotes the difference in the number of atoms of element $i$ between the system with and without the defect, for example in the case of an isolated oxygen vacancy $\Delta n_{\O}=-1$ whereas all the other $\Delta n_i$ are zero. The chemical potentials of the constituents $\mu_i$ are conveniently expressed with respect to the bulk chemical potentials $\mu_i^{\bulk}$ of the respective elemental ground states, $\mu_i=\mu_i^{\bulk}+\Delta\mu_i$. Metal and oxygen-rich conditions correspond to $\Delta\mu_\text{metal}=0$ and $\Delta\mu_{\O}=0$, respectively. The values of the chemical potentials can be translated to partial pressures enabling direct comparison with experiments. \cite{ErhAlb08} For simplicity, in the present work He-rich conditions are assumed always, {\it i.e.} $\Delta\mu_\He=0\,\eV$. For an elemental metal \Eq{eq:eform} reduces to the usual expression
\begin{align}
  \Delta E_D &= E_D - \frac{N+\Delta n}{N} E_H,
\end{align}
where $N$ denotes the number of atoms in the ideal cell.

In terms of their formation energies, the binding energy of two defects $A$ and $B$ is given by
\begin{align}
  \Delta E_b(AB) &= \Delta E_f(AB) - \Delta E_f(A) - \Delta E_f(B).
  \label{eq:ebind}
\end{align}
Following this convention negative binding energies correspond to exothermic defect reactions and imply an attractive interaction between $A$ and $B$. Note that the binding energy can change as a function of the chemical potential but is independent of the chemical environment since the $\Delta n_i$ terms that appear in \Eq{eq:eform} cancel each other in \Eq{eq:ebind}.

Assuming independent defects (low density limit) \cite{AllLid03} the equilibrium concentration of a defect is related to its free energy of formation $\Delta G_f$ according to
\begin{align}
  c_{eq} &= c_0 \exp\left[-\Delta G_f/k_B T\right],
  \label{eq:conc}
\end{align}
where the pre-factor $c_0$ denotes the density of potential defect sites per unit volume. For extrinsic defects the equilibrium concentration corresponds to the solubility of the foreign element in the matrix. The free energy of defect formation $\Delta G_f$ can be decomposed into the formation energy $\Delta E_f$ as well as the vibrational $T \Delta S_{f,vib}$ and electronic $T \Delta S_{f,el}$ formation entropies. (Note that \Eq{eq:conc} already incorporates the configurational entropy, see Ref.~\onlinecite{AllLid03}). For materials with a band gap ($E_G\gg k_BT$) the electronic contribution to the formation entropy is virtually zero and even for metals this term is usually small compared to the other contributions. The vibrational formation entropy can become important at elevated temperatures. The present work is, however, mainly concerned with comparing defects with very similar characteristics and therefore {\em relative} changes of the formation entropy between different defects can be expected to be small compared to the formation energy term. In the following, therefore only formation energies are considered and $\Delta G_f \approx \Delta E_f$.

It should be stressed that formation energies are equilibrium quantities while a material under intense irradiation is obviously a non-equilibrium system. Yet within the constraints of such a scenario, it is nonetheless useful to consider formation energies as they will determine the driving forces, which determine the long time evolution of the system.

\subsection{Computational details}

Calculations were carried out within density-functional theory using the projector augmented wave formalism \cite{Blo94, KreJou99} as implemented in the Vienna ab-initio simulation package. \cite{KreHaf93, KreHaf94, KreFur96a, KreFur96b} Ti-$3p$, Y-$4s$, Y-$4p$, Ca-$3p$, Sr-$4s$, Sr-$4p$, Ba-$5s$ as well as Ba-$5p$ states were treated as part of the valence. The plane-wave energy cutoff was set to 500\,eV for all calculations. To represent exchange and correlation effects, we employed the local density approximation (LDA), the generalized gradient approximation as parametrized by Perdew, Burke and Ernzerhof (PBE) \cite{PerBurErn96} as well as range-separated hybrid functionals \cite{HeyScuErn03} obtained by mixing conventional XC functionals (LDA or PBE) with 25\%\ exact exchange at short-range with a screening parameter of $0.2\,\text{\AA}^{-1}$. These functionals are refered to as HSE06$_\LDA$ and HSE06, respectively.

For LDA and PBE the electronic contributions to the dielectric constants reported in Table~\ref{tab:ideal_alumina} were computed within the linear response approach taking into account local field effects. For the hybrid functionals the electronic contribution was calculated from matrix elements of the dipole operator in the velocity gauge and the local field effect correction from either LDA or PBE was added. The ionic contribution was computed for LDA and PBE using linear response theory.

Details regarding the defect calculations in oxides, specifically Brillouin zone sampling, supercell cell shapes and sizes, are summarized in \tab{tab:method}. For pure iron the PBE functional was used as well as $4\times 4\times 4$ supercells containing 128 atoms and a $3\times 3\times 3$ Monkhorst-Pack grid for sampling the Brillouin zone. For charged defects the monopole-monopole correction according to Makov and Payne \cite{MakPay95} was applied using the calculated static dielectric constants given in \tab{tab:ideal_alumina}. Migration barriers were obtained via the climbing image-nudged elastic band method \cite{HenJon00, HenUbeJon00} using three intermediate images to represent the transition path. The convergence of the calculations with respect to Brillouin zone sampling and supercell size was carefully tested, based on which the error in the He interstitial formation energies due to the computational parameters is estimated to be less than 0.1\,eV.

Interfaces were modeled using slab geometries employing a similar approach as in Ref.~\onlinecite{ForWah10}. For the ideal Fe\textbar MgO interface the supercell contained six Fe and six MgO layers equivalent to 36 atoms. The Fe\textbar FeO\textbar MgO interface model was composed of four Fe, two FeO, and six MgO layers equivalent to 40 atoms. Both systems were fully relaxed including cell shape and volume until forces were below 20\,meV/\AA\ and the components of the stress tensor less than 1\,kbar. For calculations involving He interstitials the supercell size was doubled parallel to the interface leading to supercells with 144 and 160 atoms, respectively. Helium interstitial positions were systematically sampled across the interface as well as the two ``bulk'' parts corresponding to 18 distinct configurations for each interface model. Each defect configuration was relaxed at fixed cell shape and volume until the maximum force fell below 30\,meV/\AA. In these calculations the Brillouin zone was sampled using a $10\times10\times1$ Monkhorst-Pack grid for the ideal cells and a $5\times5\times1$ grid for the defect cells.

Results for bulk oxides considered in this study and an extensive comparison of different XC functionals is provided in the appendix.


\section{Helium and intrinsic defects}
\label{sect:defects}

Helium can be incorporated either substitutionally or as an interstitial defect. In the next three sections, the thermodynamics of these two forms of He in three different prototypical oxides (Al$_2$O$_3$, TiO$_2$, Y$_2$O$_3$) are compared. It is shown that interstitial He is the most stable form under most conditions and is also the most relevant form with regard to sequestration in ODS steels.

\subsection{Alumina}

\begin{figure*}
  \centering
\includegraphics[scale=0.53]{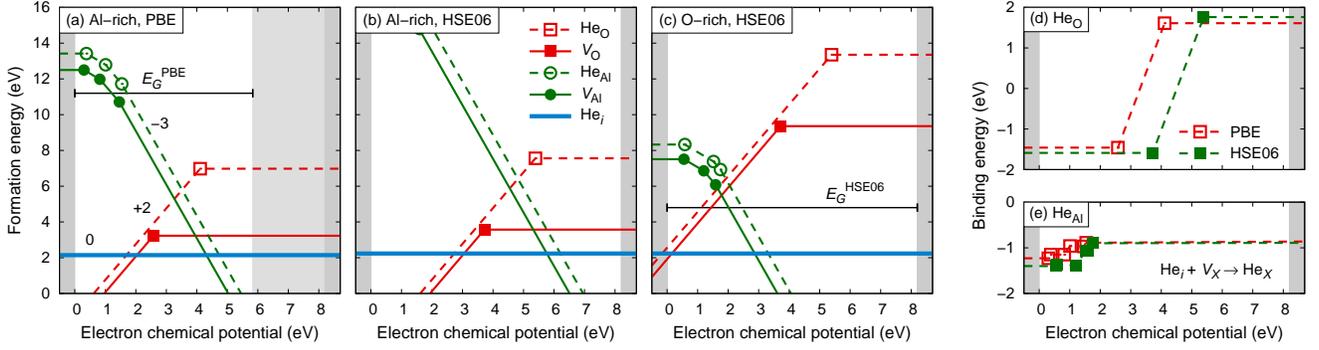}
  \caption{
    (a-c) Formation energies for point defects in alumina (a) calculated using the PBE XC functional under Al-rich conditions, and calculated using the HSE06 hybrid functional under (b) Al-rich and (c) O-rich conditions, respectively. The line slopes correspond to the defect charge states according to \Eq{eq:eform}.
    (d-e) Binding energies of He interstitials to vacancies for (d) oxygen and (e) aluminum.
  }
  \label{fig:eform_alumina}
  \label{fig:ebind_alumina}
\end{figure*}

Figure~\ref{fig:eform_alumina} shows the formation energies of interstitial and substitutional He as well as several intrinsic point defects in alumina. According to the calculations He interstitials preferentially occupy positions that are equivalent to $2b$ Wyckoff sites of the ideal structure. The interstitial formation energy is independent of the electron chemical potential, which is expected based on the closed-shell nature of He, and affected only slightly by the choice of XC functional (PBE: 2.15\,eV, HSE06: 2.23\,eV).

When He is substituted for Al the formation energy of the resulting defect closely follows the Fermi level dependence of the Al vacancy safe for a constant upward shift. In contrast, while substitutional He on an oxygen site exhibits the same charge states as the oxygen vacancy, the 2+/0 equilibrium transition level for $\He_\O$ is strongly shifted with respect to $V_\O$.

This behavior can be rationalized by considering the relaxation patterns of different defect charge states. Since He has a closed shell the formation energy difference between a vacancy and the corresponding substitutional defect results predominantly from strain. For the aluminum vacancy all charge states exhibit the same type of outward relaxation pattern and the formation energy difference between $V_\Al$ and $\He_\Al$ is only weakly affected by the charge state. In the case of the oxygen vacancy, however, the first neighbor shell relaxes inward for the neutral and outward for the positive ($2+$) charge state. This behavior is typical for ``deep'' oxygen vacancies  and also observed in other oxides including e.g., ZnO and In$_2$O$_3$ (see Ref.~\onlinecite{ErhKleAlb05,*AgoAlbNie09}). \footnote{In these materials, the oxygen vacancy also exhibits negative-$U$ type character with the $+1$ charge state being unstable with respect to the $+2$ and $0$ charge states.} As a result the strain energy contribution to the formation energies of $\He_\O$ is strongly charge state dependent, which explains the strong shift of the equilibrium transition levels from $V_\O$ to $\He_\O$ that is observable in \fig{fig:eform_alumina}.

The absolute values of vacancy and substitutional He formation energies depend on the treatment of XC. The HSE06 hybrid functional, which provides a much improved value for the band gap compared to the PBE functional (compare \tab{tab:ideal_alumina}), yields larger formation energies approximately in accord with the increase in band gap. Yet PBE and HSE06 yield qualitatively the same picture with the equilibrium transition levels of $V_\O$/$\He_\O$ and $V_\Al$/$\He_\Al$ tracking the conduction and valence band edges, respectively.

Figures~\ref{fig:eform_alumina}(b,c) illustrate not only the dependence of defect formation energies on the electron chemical potential but also on the chemical environment (compare \Eq{eq:eform} and discussion thereafter). Moving from Al-rich to O-rich conditions, which can be achieved by regulating the oxygen partial pressure, shifts the balance between $\He_\O$ ($\Delta n_\O=-1$) and $\He_\Al$ ($\Delta n_\Al=-1$) but leaves the formation energy of $\He_i$ unchanged ($\Delta n_\Al=\Delta n_\O=0$). Figure~\ref{fig:eform_alumina} shows that the formation energies of both vacancies and substitutional He can become negative under certain conditions. As negative formation energies imply the material being unstable with respect to defect formation, the Fermi level is constrained to the range in which all formation energies of intrinsic defects are positive. Considering the thermodynamically allowed regions in \fig{fig:eform_alumina}, it can be concluded that for most conditions He interstitials are thermodynamically the preferred form of He in alumina. Yet He interstitials do bind to existing vacancies, since the reaction
\begin{align*}
  \He_i + V_X \rightarrow \He_X
\end{align*}
is exothermic and the binding energy $E_b(\He_X)$ [compare \Eq{eq:ebind}] is negative as shown in \fig{fig:ebind_alumina}(d,e). The figure also demonstrates that the impact of the XC functional on the binding energies is small.

\subsection{Titania}

\begin{figure}
  \centering
\includegraphics[scale=0.53]{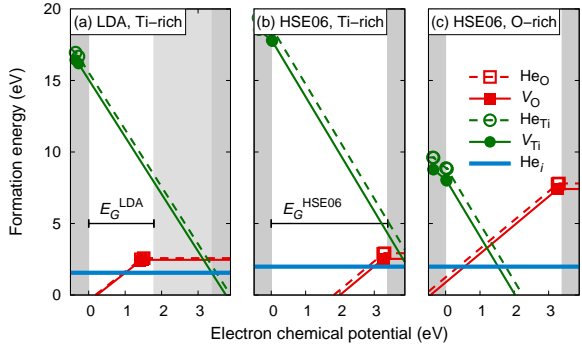}
  \caption{
    Formation energies for point defects in titania calculated using (a) the PBE XC functional under Al-rich conditions as well as the HSE06 hybrid functional under (b) Ti-rich and (c) O-rich conditions.
  }
  \label{fig:eform_titania}
\end{figure}

The formation energies of several forms of He in titania as well as related intrinsic defects are shown in \fig{fig:eform_titania}. The lowest interstitial formation energy is obtained if He occupies positions equivalent to the $4c$ Wyckoff site of the ideal structure. Similar to the case of alumina substitution on the cation sublattice leads to a defect with characteristics that are very similar to the cation vacancy. Unlike alumina this resemblance is also observed for oxygen vacancy and $\He_\O$. Once again this behavior can be related to the relaxation patterns of the oxygen vacancy. In contrast to alumina, the oxygen vacancy in titania is ``shallow'', relaxation occurs inward for all charge states, and the strain energy associated with He insertion at the vacant oxygen site is virtually independent of charge state. The binding energy between a He interstitial and a vacancy is $-1.0\,\eV$ (LDA)/$-1.1\,\eV$ (HSE06) for $\He_\Ti$ and $-1.4\,\eV$ (LDA)/$-1.6\,\eV$ (HSE06) for $\He_\O$.

The He interstitial formation energy is only weakly affected by the treatment of exchange and correlation (LDA: 1.55\,eV, HSE06: 1.98\,eV). The same applies to binding energies. As in the case of alumina He interstitials are thermodynamically the most stable form of He.

\subsection{Yttria}

\begin{figure}
  \centering
\includegraphics[width=\linewidth]{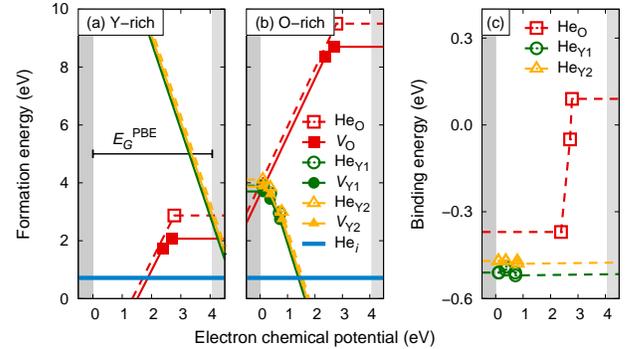}
  \caption{
    Formation energies for point defects in yttria calculated using the PBE XC functional under (a) Y-rich and (b) O-rich conditions. (c) Binding energies of He interstitials to vacancies.
  }
  \label{fig:yttria_eform}
\end{figure}

The defect formation energies for yttria shown in \fig{fig:yttria_eform} confirm the trends that were already observed for alumina and titania. The oxygen vacancy in yttria resembles its counterpart in alumina in so far as it also exhibits the characteristics of a deep defect. Helium interstitials occupy positions that are equivalent to $16c$ Wyckoff sites in the perfect lattice, which correspond to the structural vacancies of the bixbyite structure (see \sect{sect:yttria_ideal}), and they bind to vacancies.

\section{He interstitials in bulk oxides}

\begin{figure}
  \centering
\includegraphics[scale=0.6]{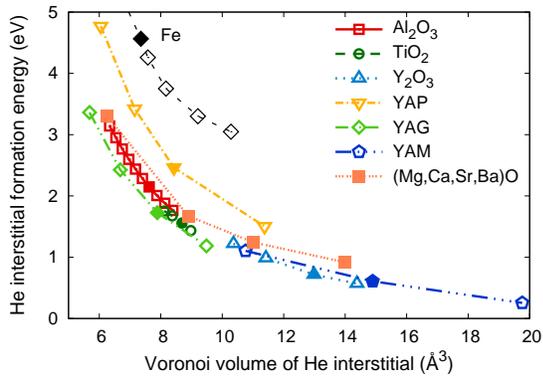}
  \caption{
    Formation energies of He interstitials in various oxides as a function of the free volume at the interstitial site. Filled symbols indicate values at the respective equilibrium volumina.
  }
  \label{fig:scaling_int}
\end{figure}

In the previous three sections it was demonstrated that within the thermodynamically allowed range of Fermi levels He interstitials prevail. The actual He interstitial formation energies for different oxides, however, vary over a wide range (Al$_2$O$_3$: 2.15\,eV, TiO$_2$: 1.55\,eV, Y$_2$O$_3$: 0.71\,eV, all values from conventional XC functionals). To resolve this variation He interstitial formation energies were computed as a function of volume for the oxides discussed above as well as the yttrium aluminum oxides and rocksalt structured oxides described in Sects.~\ref{sect:yap_ideal} and \ref{sect:mgo_ideal} of the appendix. Since the XC functional was shown in \sect{sect:defects} to have a minor influence on He interstitial formation energies, cal\-culations were carried out using conventional XC functionals that are computationally much more efficient than their hybrid relatives.

\subsection{Scaling relation for formation energies}
\label{sect:interstitials}

As shown in \fig{fig:scaling_int} the formation energies are found to scale remarkably well with the free volume at the interstitial site, where the latter is measured by the Voronoi volume of the He site in the relaxed configuration. \footnote{Voronoi volumes were obtained using the \textsc{voro++} software, see Refs.~\onlinecite{RycGreLan06, Ryc07}} It turns out that other measures such as the distance from the He site to the nearest neighbor atom (which is equivalent to constructing the smallest sphere around the He interstitial) do not yield such favorable scaling relations. Figure~\ref{fig:scaling_int} also demonstrates that the formation energies of He interstitials in iron do not fall in the range of the oxides and for the same free volumes are systematically higher.

The scaling relation in \fig{fig:scaling_int} is almost independent of the chemistry of the oxide involved and provides ample evidence that the major source of variation in the formation energies of He interstitials in oxides is the volumetric compression of the He atom. It thus effectively decouples chemistry from geometry and provides the basis for a simplified treatment of interfaces in \sect{sect:interfaces}.

\subsection{Migration barriers}
\label{sect:migbarrs}

\begin{table}
  \centering
  \caption{
    Migration barriers in eV for He interstitials in several oxides.
    Note that jump directions are approximate.
    $\Delta V_{rel}$: change in Voronoi volume of He site between initial state and saddle point normalized by volume of initial state;
    $\Delta r_{rel}$: change in He--nearest neighbor distance between initial state and saddle point normalized by initial neighbor distance.
  }
  \label{tab:migbarrs}
  \begin{tabular}{lc*{3}d}
    \hline\hline
    \multicolumn{1}{l}{Material}
    & \multicolumn{1}{c}{Direction}
    & \multicolumn{1}{c}{Barrier (eV)}
    & \multicolumn{1}{c}{$\Delta V_{rel}$ (\%)}
    & \multicolumn{1}{c}{$\Delta r_{rel}$ (\%)}
    \\
    \hline\\[-6pt]
Al$_2$O$_3$  & $11\bar{1}$ & 2.16  & -12.5  & -11.2 \\
             & $001$       & 3.86  &  -9.2  & -16.9 \\[6pt]

TiO$_2$      & $001$       & 0.11  &   3.6  &   2.5 \\
             & $110$       & 1.22  &  -9.7  &  -8.5 \\[6pt]

Y$_2$O$_3$   & $\bar{1}\bar{1}1$ & 0.70  &   1.4  & -15.0 \\
             & $111$             & 0.29  &   4.0  &  -9.4 \\
             & $0\bar{2}1$       & 2.73  &   0.0  & -23.2 \\[6pt]

YAP          & $1\bar{2}0$ & 1.22  &  24.6  &  -5.6 \\
             & $101$       & 0.98  &   3.6  &  -3.4 \\[6pt]

MgO          & $100$       & 0.76  &  16.1  & -10.2 \\
CaO          & $100$       & 0.84  &  17.2  &  -9.6 \\
SrO          & $100$       & 0.78  &  17.4  &  -9.0 \\
BaO          & $100$       & 0.61  &  17.6  &  -8.6 \\
    \hline\hline

\end{tabular}
\end{table}

\begin{figure}
  \centering
\includegraphics[scale=0.6]{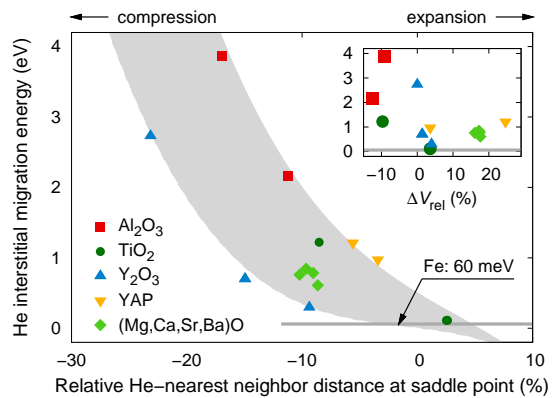}
  \caption{
    Migration barriers for He interstitials in several oxides as a function of the relative change in He--nearest neighbor separation. The inset shows the same data as a function of the relative change in the Voronoi volume of the He site. The light gray stripe is intended as a guide for the eye.
  }
  \label{fig:migbarrs}
\end{figure}

To compare the mobility of He interstitials in oxides with iron their migration barriers in several oxides were determined. For each oxide we included paths that included up to three neighbor shells on the lattice of Wyckoff sites that correspond to the most stable He position. The results are compiled in \tab{tab:migbarrs}.

To establish a scaling relation for migration barriers similar to the one for formation energies, we considered the change in free volume (again measured using the Voronoi construction) from the initial He position $V_{ini}$ to the saddle point $V_{sad}$  normalized with respect to the initial volume
\begin{align}
  \Delta V_{rel} &= \frac{V_{sad} - V_{ini}}{V_{ini}}
\end{align}
as well as the relative change in the distance between He and its nearest neighbor, $\Delta r_{rel}$, defined analogously.

Figure~\ref{fig:migbarrs} shows the migration barriers to scale fairly well with both of these measures, especially the relative He--nearest neighbor distance, but neither measure yields as convincing a scaling relation as in the case of the formation energies. The figure also contains the migration barrier for Fe, \cite{FuWil05} which at 60\,meV is substantially lower than any of the migration barriers found in the oxides. One can thus expect He interstitials in oxides to be much less mobile than in iron.

\subsection{Helium interstitial clustering}
\label{sect:clustering}

It has been established by previous first-principles calculations \cite{FuWil05} that He interstitials in Fe exhibit a strong tendency to bind both to other He interstitials and vacancies. The resulting defect complexes represent nascent He bubbles and therefore play a key role in understanding He embrittlement in iron and steels.

\begin{figure}
  \centering
\includegraphics[scale=0.6]{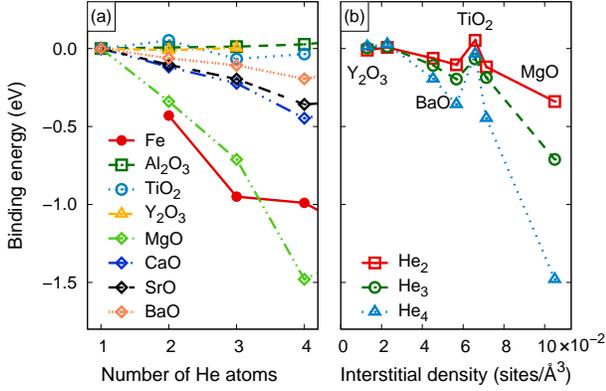}
  \caption{
    Binding energy of He interstitial clusters in several oxides as well as iron (Fe data from Ref.~\onlinecite{FuWil05}) as a function of (a) the number of He atoms in the clusters and (b) the density of He interstitial sites in the host material.
  }
  \label{fig:clustering}
\end{figure}

Complementing this information with regard to ODS steels, \fig{fig:clustering} shows binding energies of small He interstitial clusters in oxides. The left hand panel of \fig{fig:clustering} shows the dependence of the binding energy on the number of He atoms in the cluster for several oxides as well as iron. According to these data He--He interactions in oxides are typically much weaker than in iron with the exception of MgO. In some cases the interaction is even repulsive.

The trends displayed in \fig{fig:clustering}(a) can be readily explained, at least qualitatively, by considering the density of He interstitial sites in a given host and thus the average distance between nearest neighbor He interstitials. Since all He interstitials impose a relatively short-ranged compressive strain on the surrounding lattice, one can expect two He interstitials to interact weakly repulsively at moderate separations. In contrast, at very short distances two He interstitials can effectively attract each other since the total lattice strain is confined to a smaller volume and thus the total strain energy is less for a He pair than for two separate interstitials.

As shown in \fig{fig:clustering}(b) the binding energies for small He clusters do indeed exhibit a pronounced dependence on the density of interstitial sites with larger densities being associated with stronger binding. \footnote{Titania represent a special case because of its strong tetragonal anisotropy that is also apparent from e.g., the dielectric and elastic constants.} For iron, which exhibits a strong tendency to form He clusters but is not included in \fig{fig:clustering}(b), the interstitial site density is 0.53 per \AA$^3$ and thus much higher than in any of the oxides. \footnote{In body-centered cubic iron, He atoms occupy tetrahedral interstitial sites equivalent to Wyckoff sites $12d$ resulting in a density of 0.53 sites/\AA$^3$.}

The reasoning above and the data in \fig{fig:clustering} suggest that in general one can expect He clustering to be stronger in oxides with a higher He interstitial site density and that most oxides are less prone to He cluster formation than iron.


\section{Oxide-iron interfaces}
\label{sect:interfaces}

In the previous section properties of He defects in oxides have been extensively characterized, which led to the conclusion that He predominantly occurs in the form  of interstitials. Furthermore it was found that formation energies of these defects scale with the free volume at the interstitial site almost independent of the chemistry of the host oxide. Similar though less pronounced trends could also be demonstrated for migration barriers and He interstitial cluster binding energies.

The oxide particles in ODS steels vary widely in composition and structure resulting in a rich variety of oxide-matrix interfaces of seemingly arbitrary complexity. \cite{HsiFluTum10} Although some orientation relationships have been experimentally established (e.g., Fe\textbar YAM, see Ref.~\onlinecite{HsiFluTum10}), modeling these interfaces remains a computationally almost intractable task. The results of the previous section, however, suggest that it should be possible to rationalize the behavior of He at oxide-iron interfaces primarily based on the geometry of the interface and the associated free volume. Based on this rationale we decided to focus on two model systems, Fe\textbar MgO and Fe\textbar FeO\textbar MgO. These interfaces are particular suitable for a computational study since the lattice mismatch between Fe and MgO is small, which means that for the present purpose interface dislocations can be neglected.
\footnote{In general the interface dislocation network will provide additional sites for He interstitials but this effect is ignored in the present study.}

\begin{figure}
  \centering
\includegraphics[scale=0.6]{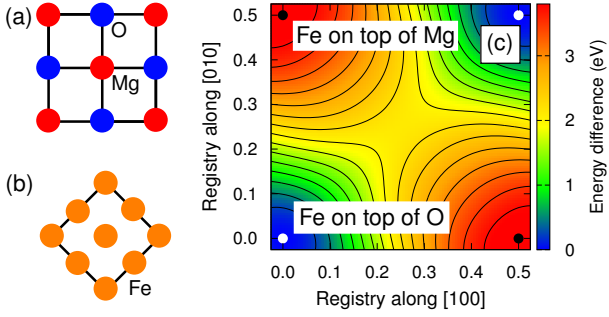}
  \caption{
    (a,b) Illustration of the Baker-Nutting orientation relationship. Projection of (a) rocksalt MgO and (b) body-centered cubic Fe along $[001]$. The latter is rotated by $45\deg$ about the [001] axis such that the $[100]_\rocksalt$ and $[110]_\bcc$ directions are parallel to each other. (c) Variation of interface energy as a function of the lateral displacement of the two crystals with respect to each other.
  }
  \label{fig:bakernutting}
\end{figure}

\subsection{Ideal interfaces}

The interface between two materials with rocksalt and body-centered cubic structure, respectively, is described by the Baker-Nutting orientation relationship. \cite{ForWah10} It is illustrated in \fig{fig:bakernutting}, which shows that the $(001)_\rocksalt$ and $(001)_\bcc$ planes as well as the $[100]_\rocksalt$ and $[110]_{\bcc}$ directions are parallel to each other. Among the earth alkaline oxides MgO has the smallest lattice mismatch with Fe (4\%\ based on the experimental lattice constants) and was therefore selected for the present study. Incidentally Fe--MgO interfaces have recently attracted a lot of attention since they exhibit a strong transverse magnetic resistance effect. \cite{LucBenLib05, FenBenAlo09, NakAkiIto10, YanChsKal10, WanZhaZha10}

\begin{figure}
  \centering
\includegraphics[scale=0.62]{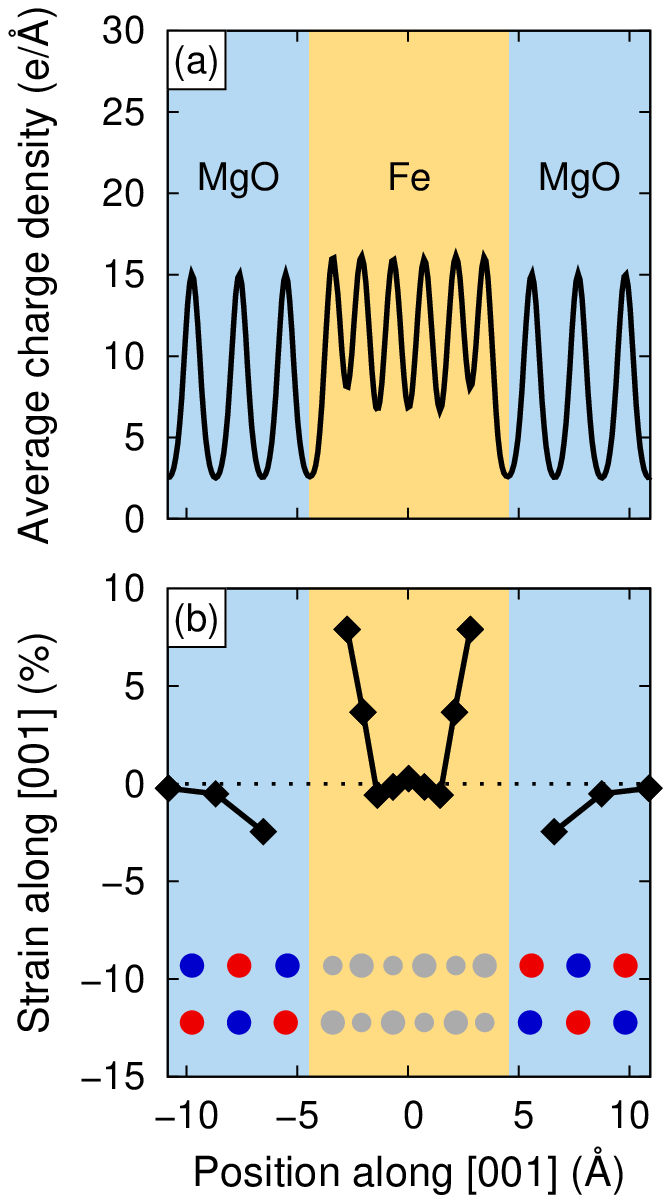}
\includegraphics[scale=0.62]{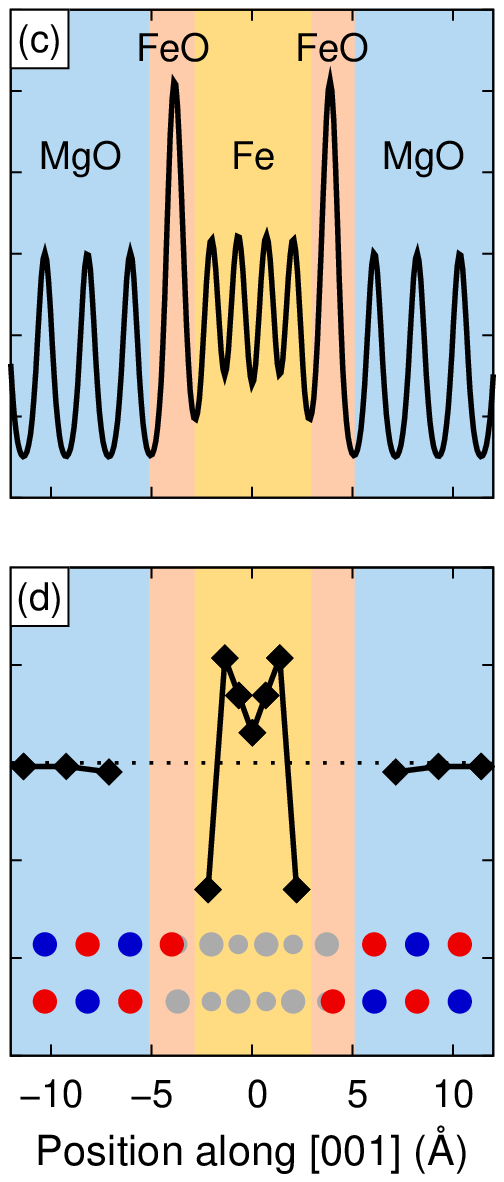}
  \caption{
    Plane-averaged charge density and out-of-plane strain for the (a,b) Fe\textbar MgO and (c,d) Fe\textbar FeO\textbar MgO interfaces as a function of position perpendicular to the interface plane. The colored spheres in the bottom panel indicate the atomic positions.
  }
  \label{fig:interface_geometry}
\end{figure}

In the present work both Fe\textbar MgO and Fe\textbar FeO\textbar MgO interface models were included, the geometries of which are illustrated in \fig{fig:interface_geometry}. First the minimum energy configuration of the Fe\textbar MgO interface was established by scanning the energy landscape as a function of in-plane displacement, which yields the Fe-on-top-of-O configuration as the most stable one. The Fe\textbar FeO\textbar MgO model is then obtained from the Fe\textbar MgO model by inserting O atoms in the outermost Fe layers such that the Fe and O atoms form a square lattice parallel to the interface. Both models were subsequently fully relaxed allowing both ionic motion as well as cell shape and volume changes.

After relaxation of the Fe\textbar MgO interface model the Fe slab is under a compressive in-plane strain of $4.8\%$ while the MgO half is under a tensile in-plane strain of $-1.2\%$. The strain along [001] is of the respective opposite sign near the interface as shown in \fig{fig:interface_geometry}(b) and quickly decays to the bulk value with increasing distance to the interface. For the Fe\textbar FeO\textbar MgO geometry the in-plane strains for the Fe and MgO part are 5.2\%\ and approximately zero, respectively. Distinctly non-zero strain along [001] is only observed for the Fe slab, for which the strain is tensile directly next to the FeO layer but compressive anywhere else. The different magnitudes of the strain in the Fe and MgO parts reflect the fact that the tetragonal shear modulus is almost 1.5 times higher for MgO than for Fe.


\subsection{Helium at interfaces}

\begin{figure}
  \centering
\includegraphics[scale=0.62]{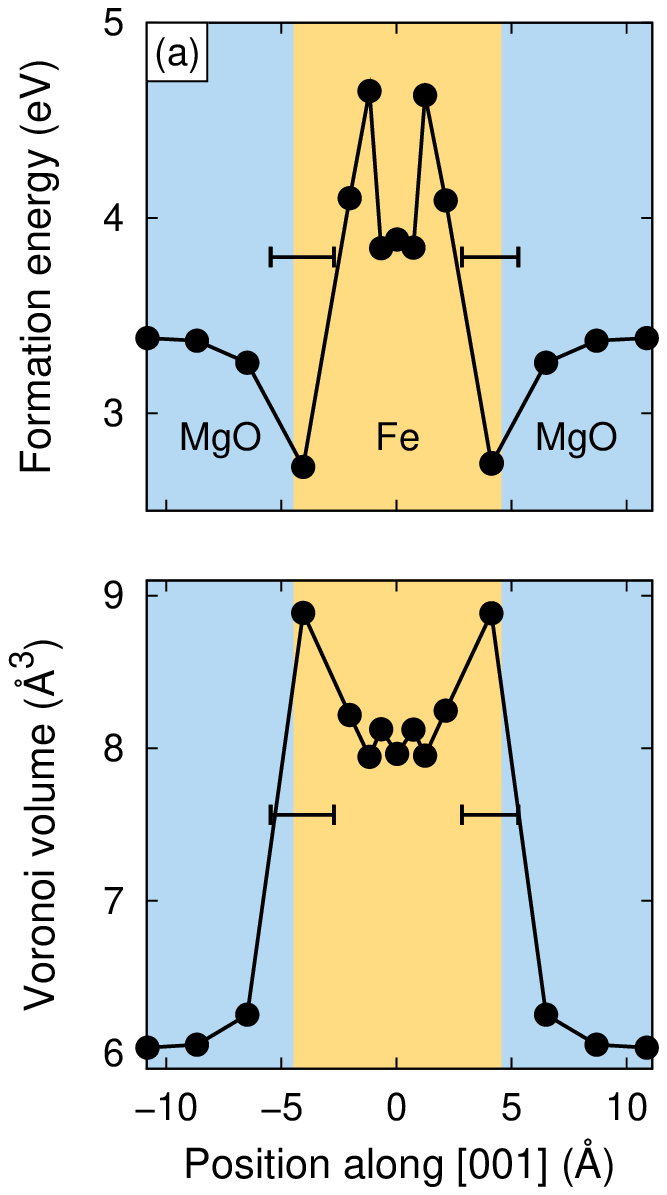}
\includegraphics[scale=0.62]{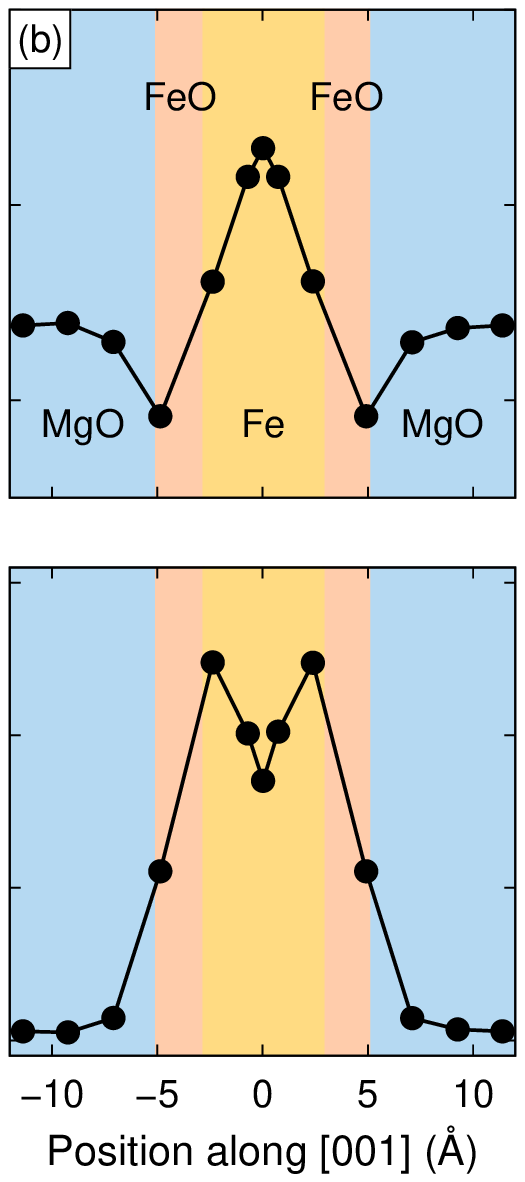}
  \caption{
    Helium interstitial formation energies and their respective Voronoi volumina for the (a) Fe\textbar MgO and (b) Fe\textbar FeO\textbar MgO interfaces as a function of position perpendicular to the interface. Helium atoms placed in the range indicated by the horizontal bars relax into the interface.
  }
  \label{fig:interface_eform}
\end{figure}

Starting from the fully relaxed interface models, He interstitials were inserted sampling all distinct known bulk sites as well as sites in the interface. The calculated formation energies for these configurations are shown in \fig{fig:interface_eform}.

For both interface configurations the formation energy for He interstitials is by far the lowest if the latter are located at the interface. For example in the case of the Fe\textbar MgO interface the formation energy is 2.7\,eV to be compared with values of 3.3\,eV and 3.9\,eV near the center of the MgO and Fe slabs, respectively. In the MgO part the formation energy is thus almost identical to the bulk value at the equilibrium lattice constant, whereas the corresponding value for Fe is noticeably lower than its unstrained bulk counterpart. This behavior is caused by two effects:
Firstly, the tetragonal shear modulus is softer for iron than for the oxide, as a result of which the Fe slab is more severely strained. Secondly, the Poisson ratio of iron is less than 0.29, which implies that the average volume per atom changes with strain. In the case at hand, one observes an increase in the free volume at the He interstitial site from 7.4\,\AA$^3$ to about 8.0\,\AA$^3$, which in combination with \fig{fig:scaling_int} readily explains the observed decrease in the formation energy.\footnote{
  The direct effect of tetragonal strain on the formation energies of He interstitials in iron was tested by considering volume conserving deformation (effectively treating the material as if it had a Poisson ratio of 0.5). For a linear strain of 5\%\ the formation energy and free volume change by only 0.16\,eV and 1.5\%\, respectively.}

Comparison of formation energies and Voronoi volumina shows that inside both the Fe and MgO slabs the formation energy scales with the free volume. At the Fe\textbar MgO interface the lowest formation energy also corresponds to the largest free volume. In the case of the Fe\textbar FeO\textbar MgO interface the formation energy is lower at the FeO\textbar MgO interface than at the FeO\textbar Fe interface even though the free volumes suggest the opposite trend. This effect is related to the earlier observation that Fe does not fall on the same scaling relation as the oxides (see \fig{fig:scaling_int}). While this implies that one cannot predict formation energies at interfaces based {\em exclusively} on free volume, it nonetheless demonstrates clearly that He interstitials are more strongly attracted to interfaces than to either one of the bulk phases. For the two systems considered here as well as e.g., Fe\textbar Y$_2$O$_3$ interfaces, \cite{Erh_unpublished} one observes an increase in the free volume at the interface compared to the bulk phases. This behavior can be attributed to a comparably weak adhesion between the oxide and the iron matrix, which is related to the transition from mixed ionic-covalent to metallic bonding across the interface.


\section{Discussion and conclusions}
\label{sect:discussion}

In this section the implications of the foregoing investigation for the understanding of He sequestration in ODS steels will be discussed. To simplify the discussion, first the major results of this work will be recapped.

A detailed examination of three different oxides of different chemical composition, stoichiometry and structure using both conventional as well as hybrid XC functionals showed that intrinsic defects limit the range within which the electron chemical potential can vary. Within the thermodynamically allowed range He interstitials are the most stable form of helium, yet they will bind to existing vacancies. While conventional and hybrid functionals yield different values for band gaps as well as absolute defect formation energies for vacancies and substitutional He, the results are qualitatively similar. Furthermore, for He interstitials conventional and hybrid functionals provide formation energies that are also quantitatively similar. Since the latter are computationally much more demanding, further calculations employed conventional functionals only.

Compared to iron He interstitial formation energies in oxides are significantly lower and exhibit a considerable materials dependence. Using data for a wide range of different oxides and volumes it was shown that the latter dependence can be described with good accuracy by a scaling relation based on the Voronoi volume of the He site. This finding demonstrates that the behavior of He can be largely rationalized in terms of free volume. It thereby greatly simplifies the task of understanding the behavior of He in ODS steels since it allows us to separate structure and chemistry.

He interstitial migration barriers were found to be systematically higher than in Fe. While the migration energies do not obey a scaling relation as cleanly as for the formation energies, as a general trend the migration barriers decrease if the relative change in the He--nearest neighbor distance from the initial configuration to the saddle point increases.

In iron He interstitials bind strongly to each other, which facilitates the formation of He bubble nuclei. \cite{FuWil05} In contrast in most oxides the binding energies between He interstitials are much smaller indicating a weaker propensity for bubble formation. Binding between He interstitials was observed to scale with the volume density of He interstitial sites, with higher densities (e.g., MgO, Fe) leading to stronger binding.

The variability and complexity of oxide-iron interfaces in ODS steels prevents a direct first-principles study of He sequestration at these interfaces. In the present study it was therefore decided to study two particular simple interfaces, Fe\textbar MgO and Fe\textbar FeO\textbar MgO. In view of the scaling relations described above, one can expect these interfaces to act as prototypes for the types of interfaces that occur in real materials. The calculations revealed that in both interfaces the formation energies of He interstitials are significantly lower than in either of the bulk phases. The low formation energies could again be correlated with a larger free volume at the interfaces. While in the MgO slabs the He interstitial formation energies reached almost exactly the value that was obtained earlier for bulk MgO, formation energies for positions inside the Fe slabs deviated significantly from the bulk value. This behavior is directly related to strain fields that distort the Fe lattice and thereby affect the free volume available at interstitial sites. The strain effect is much larger in the Fe matrix because its tetragonal shear modulus is considerably lower than the one of MgO.

\begin{figure}
  \centering
\includegraphics[width=0.98\columnwidth]{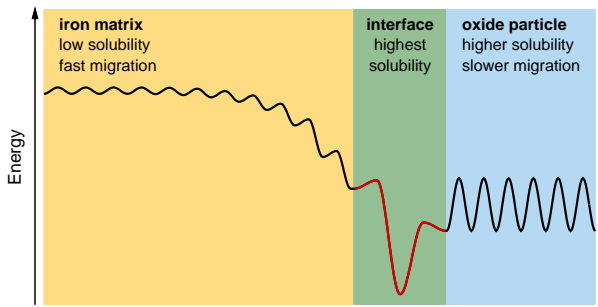}
  \caption{
    Schematic energy landscape for He interstitial migration in an ODS steel. In the Fe matrix formation energies are high but migration barriers are low, while the opposite applies for the oxide particles. The smallest formation energies and thus the highest solubilities are predicted in the interface region. Strain fields can lead to gradients near the interface that depending on the sign of the strain field can either increase or decrease toward the interface.
  }
  \label{fig:schematic}
\end{figure}

Now one can combine all this information to develop a schematic energy landscape for He interstitials in ODS steels. In the Fe matrix He interstitial formation energies are high but migration barriers are low. In contrast, formation energies in the oxide particles are lower while migration barriers are higher than in Fe. The lowest formation energies are observed at the interface. Since iron is elastically softer than the majority of oxides considered in this study, the iron matrix is more likely to be strained. This affects the free volume available to He interstitials and accordingly their formation energies (and solubilities). Combining these data, one can obtain a schematic energy landscape as the one sketched in \fig{fig:schematic}. In this particular plot the free volume in the Fe matrix is assumed to increase toward the interface leading to a gradual sloping of the landscape toward the oxide particle. It is, however, equally possible that the free volume decreases toward the interface, leading to a landscape that rises toward the interface. In reality the character of the strain field sensitively depends on the structure of the interface, which can include e.g., interface dislocations, amorphous regions, and chemical gradients (in so far as they translate to strain). In fact, experimentally He bubble formation is observed not to occur around all precipitates, \cite{HsiFluTum10} for which strain effects could be a reasonable explanation.

In the introduction, it was pointed out that oxide particles in ODS steels exhibit a broad spectrum of structural and chemical variations, including amorphous regions and extremely small nanometer-sized inclusions. While these structures were not explicitly studied here, the present work has established clear trends that hold for a variety of different local environments both structurally and chemically. It was also demonstrated that He incorporation in oxides is qualitatively different from iron. This is most clearly visible in the He interstitial formation energies (\fig{fig:scaling_int}), for which oxides display systematically lower values than iron for the same free volumes. This behavior can be readily understood in terms of mixed covalent--ionic bonding in the oxides {\it vs} metallic bonding in iron. The covalent character of oxides leads to stronger directional bonding and a more localized electron charge density compared to iron. As a result oxides adopt more open structures with larger interstitial sites (both in ionic and electronic terms), and defect induced strain fields are less extended than in metals. These qualitative features are also found in amorphous oxide particles and small oxide inclusions. (As shown in Refs.~\onlinecite{FuKrcPai07} and \onlinecite{JiaSmiOde09}, already very small oxide clusters containing just a few atoms feature pronounced directional bonding). It is therefore expected that the results obtained in the present work also transpire to these more general situations.

In summary, in this paper it has been demonstrated that oxide particles in ODS steels have a higher He solubility than the Fe matrix, which is primarily the result of larger interstitial sites. The solubility at oxide-iron interfaces is even larger than in the bulk oxides. Strain fields, which can affect in particular the iron matrix, lead to solubility gradients near oxide-iron interfaces. The data obtained in this study not only provides valuable insight into the behavior of He in ODS steels but can be used to derive parameters for rate equation models of He sequestration in ODS steels. \cite{MarBul11}


\begin{acknowledgments}
I would like to thank J. Marian for his continuous interest in this work and many fruitful discussions. Helpful discussions of experimental findings with L. Hsiung and M. Fluss are gratefully acknowledged. This work has been performed under the auspices of the U.S. Department of Energy by Lawrence Livermore National Laboratory under Contract DE-AC52-07NA27344 with support from the Laboratory Directed Research and Development Program.
\end{acknowledgments}


\appendix*

\section{Bulk properties of oxides}
\label{sect:oxides}

This appendix summarizes results for groundstate properties of the oxides included in this study as obtained using different computational methods. Based on these results the parameters for the defect calculations were chosen, in particular the XC functionals.

\subsection{Alumina}
\begin{table}
  \newcommand{\spr}[1]{\multicolumn{1}{c}{#1}}
  \centering
  \caption{
    Structural and electronic properties of Al$_2$O$_3$, TiO$_2$, and Y$_2$O$_3$ from experiment and calculations.
    $a$, $c$: lattice parameters (\AA),
    $\theta$: rhombohedral angle (deg),
    $x_i$, $y_i$, $z_i$, $u$: internal structural parameters,
    $E_G$: band gap (eV),
    $epsilon_{\infty}$: electronic contribution to dielectric constant,
    $epsilon_{\ion}$: ionic contribution to dielectric constant,
    $epsilon_{0}$: static dielectric constant (sum of $\epsilon_{\infty}$ and $\epsilon_{\ion}$).
    The subscripts $\perp$ and $||$ indicate the dielectric constant perpendicular and parallel to the rhombohedral [111] axis (equivalent to the [0001] axis in the hexagonal setting), respectively. Note that for alumina the thermal band gap is reduced with respect to the optical gap (given here) due to polaronic effects (Ref.~\onlinecite{Fre90}). Experimental data for alumina from Refs.~\onlinecite{IshMiyMin80, Fre90}, for titania from Refs.~\onlinecite{BurHugMil87, Cro52, Par61}, and for yttria from Refs.~\onlinecite{BalBerCal98, OhtYamMiy04}.
  }
%
  \label{tab:ideal_alumina}
  \label{tab:ideal_titanina}
  \label{tab:ideal_yttria}
  \begin{tabular}{l*{5}d}
    \hline\hline
    \multicolumn{6}{c}{
      Alumina (corundum, Al$_2$O$_3$), space group R$\bar{3}c$, number 167} \\
    & \spr{Expt.}
    & \spr{LDA}
    & \spr{PBE}
    & \spr{HSE06$_{\text{LDA}}$}
    & \spr{HSE06}
    \\
    \hline
    $a$                         &  4.754          &  4.737  &  4.811  &  4.712  &  4.754  \\
    $c$                         & 12.99           & 12.910  & 13.126  & 12.855  & 12.982  \\
    $x_{\Al}$                   &  0.3523         &  0.3520 &  0.3522 &  0.3519 &  0.3519 \\
    $x_{\O}$                    &  0.3064         &  0.3056 &  0.3061 &  0.3061 &  0.3063 \\[3pt]
    $E_G$                       &  8.8            & 6.38    & 5.83    & 8.59    & 8.20    \\
    $\epsilon^{\infty}_{\perp}$ & \spr{2.9-3.2}   & 3.28    & 3.27    & 2.79    & 2.79    \\
    $\epsilon^{\infty}_{||}$    &                 & 3.23    & 3.22    & 2.76    & 2.77    \\
    $\epsilon^{\ion}_{\perp}$   &                 & 5.9     & 6.8     &         &         \\
    $\epsilon^{\ion}_{||}$      &                 & 7.8     & 9.2     &         &         \\
    $\epsilon^{0}_{\perp}$      & \spr{9.3-9.5}   & 9.2     & 10.0    & 8.7     & 9.6     \\
    $\epsilon^{0}_{||}$         & \spr{11.5-11.6} & 11.1    & 12.4    & 10.6    & 12.0    \\
    \\[-3pt]
    \hline\hline
    \multicolumn{6}{c}{
      Titania (rutile, TiO$_2$), space group P$4_2/mnm$, number 136} \\
    & \spr{Expt.}
    & \spr{LDA}
    & \spr{PBE}
    & \spr{HSE06$_{\text{LDA}}$}
    & \spr{HSE06}
    \\
    \hline
    $a$                      & 4.587            & 4.557  & 4.653  & 4.542  & 4.592  \\
    $c$                      & 2.954            & 2.924  & 2.970  & 2.922  & 2.948  \\
    $u$                      & 0.3047           & 0.3038 & 0.3049 & 0.3046 & 0.3051 \\[3pt]
    $E_G$                    & 3.05             & 1.79   & 1.77   & 3.37   & 3.39   \\
    $\epsilon^{\infty}_{\perp}$  & 6.84             & 7.95   &        & 4.68   & 4.56   \\
    $\epsilon^{\infty}_{||}$    & 8.43             & 9.39   &        & 5.31   & 5.32   \\
    $\epsilon^{\ion}_{\perp}$   &                  & 164.7  &        &        &        \\
    $\epsilon^{\ion}_{||}$     &                  & 285.2  &        &        &        \\
    $\epsilon^{0}_{\perp}$     & \spr{111, 86}    & 172.6  &        &        &        \\
    $\epsilon^{0}_{||}$        & \spr{257, 170}   & 294.6  &        &        &        \\
    \\[-3pt]
    \hline\hline
    \multicolumn{6}{c}{
      Yttria (bixbyite, Y$_2$O$_3$), space group I$a\bar{3}$, number 206} \\
    & \spr{Expt.}
    & \spr{LDA}
    & \spr{PBE}
    & \spr{HSE06}
    \\
    \hline
    $a$                 & 10.5961  & 10.5115 & 10.71   & & 10.64   \\
    $x_\Y$              & -0.0326  & -0.0329 & -0.0328 & & -0.0329 \\
    $x_\O$              &  0.3908  &  0.3907 &  0.3908 & &  0.3907 \\
    $y_\O$              &  0.1519  &  0.1517 &  0.1518 & &  0.1517 \\
    $z_\O$              &  0.3801  &  0.3799 &  0.3800 & &  0.3800 \\[3pt]
    $E_G$               &  6.0     &  4.10   &  4.07   & &  5.689  \\
    $\epsilon^{\infty}$ &          &         &  4.04   & &  3.33   \\
    $\epsilon^{\ion}$   &          &         & 10.3    & &         \\
    $\epsilon^{0}$      &          &         & 14.3    & &         \\
    \hline\hline
  \end{tabular}
\end{table}

The ground state structure of alumina (Al$_2$O$_3$) is corundum ({\it Strukturbericht} symbol D5$_1$, space group R$\bar{3}c$, number 167), which has a primitive unit cell of rhombohedral symmetry containing ten atoms with Al and O atoms occupying Wyckoff sites $4c$ and $6e$, respectively. The structure can also be described using a hexagonal setting, in which case the unit cell compromises three times as many atoms. The relation between the two settings is discussed in detail in Ref.~\onlinecite{Cou81}. Following common procedure the structural parameters in \tab{tab:ideal_alumina} are given in the hexagonal setting. The data shows that LDA calculations provide slightly better agreement with structural parameters than PBE calculations. The band gap underestimation in these calculations is typical for conventional XC functionals. This deficiency is largely corrected by using hybrid functionals, which also further improve the structural parameters. For defect calculations the PBE and HSE06 functionals were used.

\subsection{Titania}
There is a wide range of titanium oxides with different stoichiometries and crystal structures. Here the rutile structure ({\it Strukturbericht} symbol C4, space group P$4_2/mnm$, number 136) is considered, which has a primitive unit cell of tetragonal symmetry containing 2 Ti and 4 O atoms, which occupy Wyckoff sites $2a$ and $4f$, respectively. For the structural parameters the closest agreement with experiment is obtained for the LDA and HSE06 functionals, which were accordingly used in defect calculations.

The PBE functional yields extremely soft phonon modes at $\Gamma$, which leads to a numerical divergence in the ionic dielectric constant. In reality TiO$_2$ does indeed possess soft modes that give rise to a very large dielectric constant. Due to their softness it is not trivial to accurately reproduce their frequencies and more crucially the resulting ionic contribution to the dielectric constant. In fact, the PBE functional did not only caused problems in calculations of the dielectric response but also in a number of defect calculations involving He interstitials: After removing the He interstitial from the relaxed supercell, the configuration did not relax to the original lattice positions but instead assumed a distorted structure with an energy that was {\em lower} than for the perfect rutile lattice. These problems raise concerns regarding the reliability of the PBE functional for calculations in TiO$_2$. These problems were specific for the PBE functional and did not transpire to the HSE06 hybrid functional.

\subsection{Yttria}
\label{sect:yttria_ideal}
The ground state structure of yttria ($\Y_2\O_3$) is bixbyite ({\it Strukturbericht} symbol D5$_3$, space group I$a\bar{3}$, number 206). The primitive cell contains 16 Y atoms on Wyckoff sites $8b$ and $24d$ as well as 24 O atoms on Wyckoff sites $48e$. The bixbyite structure can be described as a calcium fluorite lattice in which one quarter of the anion sites are unoccupied (``structural vacancies''). Structural and selected electronic properties obtained using different XC approximations are shown in \tab{tab:ideal_yttria} in comparison with experimental data demonstrating good overall agreement. For this material defect calculations were carried out using the PBE XC functional only.

\subsection{Yttrium aluminum oxides}
\label{sect:yap_ideal}
\begin{table}
  \newcommand{\spr}[1]{\multicolumn{1}{c}{#1}}
  \centering
  \caption{
    Structural and electronic properties of yttrium aluminum oxides. Symbols as in \tab{tab:ideal_alumina}.
    Experimental data for YAP, YAG and YAM from Refs.~\onlinecite{Ros96}, \onlinecite{NakYosYam99}, and \onlinecite{YamOmoHir95}, respectively.
  }
  \label{tab:ideal_yap}
  \label{tab:ideal_yag}
  \label{tab:ideal_yam}
  \begin{tabular}{l*{3}d}
    \hline\hline
    \multicolumn{4}{c}{
      YAP (YAlO$_3$), space group P$nma$, number 62} \\
    & \spr{Experiment}
    & \spr{LDA}
    & \spr{PBE}
    \\
    \hline
    $a$        & 5.315  & 5.284  & 5.392 \\
    $b$        & 7.354  & 7.314  & 7.452 \\
    $c$        & 5.167  & 5.133  & 5.226 \\
    $x_{\Y1}$  & 0.053  & 0.054  & 0.055 \\
    $z_{\Y1}$  & 0.988  & 0.987  & 0.987 \\
    $x_{\O1}$  & 0.477  & 0.480  & 0.477 \\
    $z_{\O1}$  & 0.083  & 0.083  & 0.086 \\
    $x_{\O2}$  & 0.292  & 0.293  & 0.294 \\
    $y_{\O2}$  & 0.042  & 0.044  & 0.045 \\
    $z_{\O2}$  & 0.707  & 0.706  & 0.706 \\
    \\[-3pt]
    \hline\hline
    \multicolumn{4}{c}{
      YAG (Y$_3$Al$_5$O$_{12}$), space group I$a\bar{3}d$, number 230} \\
    & \spr{Experiment}
    & \spr{LDA}
    & \spr{PBE}
    \\
    \hline
    $a$       & 12.006 & 11.916 & 12.127 \\
    $x_{\O}$  & -0.032 & -0.031 & -0.031 \\
    $y_{\O}$  &  0.051 &  0.051 &  0.050 \\
    $z_{\O}$  &  0.150 &  0.149 &  0.149 \\
    \\[-3pt]
    \hline\hline
    \multicolumn{4}{c}{
      YAM (Y$_4$Al$_2$O$_9$), space group P$2_1/c$, number 14} \\
    & \spr{Experiment}
    & \spr{LDA}
    & \spr{PBE}
    \\
    \hline
    $a$       &   7.375 &   7.296 &   7.441 \\
    $b$       &  10.462 &  10.380 &  10.563 \\
    $c$       &  11.110 &  11.044 &  11.260 \\
    $\beta$   & 108.58  & 108.46  & 108.50  \\[3pt]
    \hline\hline
  \end{tabular}
\end{table}
There are three distinct ground state structures in the Y--Al--O system that contain all three elements, a perovskite (yttrium aluminate, YAP), a garnet (YAG), and a monoclinic (YAM) structure. These ternary oxides are included because yttrium and aluminum oxide are common additions in ODS steels and some of the resulting oxides have been observed experimentally. \cite{HsiFluTum10}

The perovskite structure (YAP) has orthorhombic symmetry with space group 62 (P$nma$). \footnote{
  In the international crystallographic database some records for this structure are provided in the P$bnm$ setting. The P$nma$ and P$bnm$ settings are equivalent and can be easily transformed into each by swapping the axes. The Bilbao crystallographic server (\texttt{http://www.cryst.ehu.es}) uses the P$nma$ setting.
}
Its primitive unit cell contains 4 Al, 4 Y, and 12 O atoms corresponding to a chemical sum formula of YAlO$_3$. Yttrium occupies $4c$ Wyckoff sites with two internal parameters, $x_{\Y}$ and $z_{Y}$ while the Al atoms sit on the $4b$ Wyckoff sites with no internal parameters. The oxygen atoms occupy both $4c$ and $8d$ sites. Their positions can be described using five internal parameters, $x_{\O1}$, $z_{\O1}$, $x_{\O2}$, $y_{\O2}$, and $z_{\O2}$. \tab{tab:ideal_yap} compares the structural data obtained from experiments and calculations.

The garnet structure (YAG) has cubic symmetry with space group 230 (I$a\bar{3}d$). Its primitive unit cell contains 20 Al, 12 Y, and 48 O atoms corresponding to a chemical sum formula of Y$_3$Al$_5$O$_{12}$. Yttrium occupies the $24c$ Wyckoff sites and aluminum atoms are located both on $16a$ and $24d$ Wyckoff sites. Oxygen atoms occupy the $96h$ Wyckoff sites with three internal parameters, $x_\O$, $y_\O$, and $z_\O$. The structural data obtained from experiments and calculations are compared in \tab{tab:ideal_yag}.

Monoclinic yttrium aluminum oxide (YAM) belongs to space group 14 (P$2_1/c$). Its primitive unit cell contains 8 Al, 16 Y, and 36 O atoms corresponding to a chemical sum formula of Y$_4$Al$_2$O$_9$. Yttrium, aluminum, and oxygen occupy 4, 2, and 9 different $4e$ Wyckoff sites, respectively, leading to a total of 45 internal parameters. \tab{tab:ideal_yam} compares the lattice constants to experimental data revealing good overall agreement with the reference data. We refrain from listing the internal structural parameters but assert that they are also in good agreement with experiment.

\subsection{Oxides in the rocksalt structure}
\label{sect:mgo_ideal}
There are a number of oxides that adopt the rocksalt structure. Here only the earth alkaline oxides are included as the most simple variant. One obtains lattice constants of 4.240\,\AA\ (experimental value 4.207\,\AA, Ref.~\onlinecite{Landolt}), 4.832\,\AA\ (4.803\,\AA, Ref.~\onlinecite{Landolt}), 5.207\,\AA\ (5.160\,\AA, Ref.~\onlinecite{Wyc63}), and 5.611\,\AA\ (5.524\,\AA, Ref.~\onlinecite{Zol55}) for MgO, CaO, SrO, and BaO, respectively.

\end{document}